\documentclass[12pt,a4j]{article}
\usepackage{amsmath,amsthm,amssymb,bm,fancybox,multirow,hhline,psfrag,comment}
\usepackage{url}
\usepackage{color}
\usepackage[dvipdfm]{graphicx}
\usepackage{natbib}

\def\hsymbu#1{\smash{\lower1.7ex\hbox{\huge$#1$}}}

\usepackage{indentfirst}
\usepackage{algorithm,algorithmic}

\bibpunct{(}{)}{;}{a}{}{,}

\def\E{{E}}

\def\l{{l}}

\makeatletter
\def\eqnarray{\stepcounter {equation}\let \@currentlabel =\theequation
\global \@eqnswtrue
\global \@eqcnt \z@ \tabskip \@centering \let \\=\@eqncr
$$\halign to \displaywidth \bgroup \@eqnsel \hskip \@centering
$\displaystyle \tabskip \z@ {##}$&\global \@eqcnt \@ne \hfil
${\mbox{}##\mbox{}}$\hfil &\global \@eqcnt \tw@
$\displaystyle \tabskip \z@ {##}$\hfil \tabskip \@centering
&\llap {##}\tabskip \z@ \cr}
\makeatother

\begin{document}

\theoremstyle{plain}
\newtheorem{lemma}{Lemma}[section]
\theoremstyle{remark}
\newtheorem{remark}{Remark}[section]
\theoremstyle{example}
\newtheorem{example}{Example}[section]

{
\begin{center}
\textbf{\Large Full information maximum likelihood estimation in factor analysis with a lot of missing values}
\end{center}
\begin{center}
\large {Kei Hirose, Sunyong Kim, Yutaka Kano, Miyuki Imada, Manabu Yoshida and Masato Matsuo }
\end{center}

\begin{center}
{\it {\small
Division of Mathematical Science, Graduate School of Engineering Science, Osaka University,\\
1-3, Machikaneyama-cho, Toyonaka, Osaka, 560-8531, Japan \\

\vspace{1.2mm}

$^2$ NTT Network Innovation Laboratories, 1--1, Hikarinooka, Yokosuka-shi, Kanagawa, 239--0847 Japan.\\

$^3$ NTT Network Innovation Laboratories, 3--9--11, Midori-cho, Musashino-shi, Tokyo, 180--8585 Japan.\\

}}
{\it {\small E-mail: hirose@sigmath.es.osaka-u.ac.jp, 
}}
\end{center}
\begin{abstract}
We consider the problem of full information maximum likelihood (FIML) estimation in a factor analysis model when a majority of the data values are missing.  
The expectation--maximization (EM) algorithm is often used to find the FIML estimates, in which the missing values on observed variables are included in complete data.  
However, the EM algorithm has an extremely high computational cost when the number of observations is large and/or plenty of missing values are involved.  
In this paper, we propose a new algorithm that is based on the EM algorithm but that efficiently computes the FIML estimates.  
A significant improvement in the computational speed is realized by not treating the missing values on observed variables as a part of complete data.  
Our algorithm is applied to a real data set collected from a Web questionnaire that asks about first impressions of human; almost 90\% of the data values are missing.  
When there are many missing data values, it is not clear if the FIML procedure can achieve good estimation accuracy even if the number of observations is large.  
In order to investigate this, we conduct Monte Carlo simulations under a wide variety of sample sizes. 
\end{abstract}

\section{Introduction}
Factor analysis provides a practical tool for exploring the covariance structure among a set of observed random variables by constructing a smaller number of unobserved variables called common factors.  Successful applications have been reported in various fields of research, including the social and behavioral sciences.   In practical situations, a majority of the data values are often missing or unknown.  For example, when a questionnaire asks a research participant about a feeling toward another person, a number of questions are needed to investigate their impressions, using a wide variety of personal-assessment measures.  However, answering all of the questions may cause participants fatigue and inattention, resulting in inaccurate answers.  In order to gather the high-quality data, the participants may be asked to select just a few of questions; this leads to a large number of missing values. 

In the presence of missing values, the factor analysis model can be estimated by the full information maximum likelihood (FIML) procedure.  It is well-known that the FIML method yields a consistent estimator under the assumption of missing at random (MAR, e.g., \citealp{little1987statistical}), that is, the missingness depends only on the variables that are observed and not on the missing values.

There are two crucial issues for the FIML procedure with large rates of missing values. The first issue is the computational speed.  Conventionally, FIML estimates have been obtained by Newton-type algorithms.  For example, \citet{finkbeiner1979estimation} applied a quasi-Newton method to the factor analysis model, and \citet{lee1986estimation} considered an estimation of the general covariance structure via the reweighed Gauss--Newton algorithm.  However, Newton-type methods can be slow and unstable when the number of variables is large. Another popular estimation algorithm is the expectation--maximization (EM) algorithm and its extensions; in this approach, the common factors and missing values on observed variables are included in complete data (e.g., \citealp{dempster1977maximum,rubin1982algorithms,little1987statistical,jamshidian1993conjugate,jamshidian1997algorithm,liu1998maximum}). However, the ordinary EM algorithm also has a high computational cost when a majority of the data values are missing, because a number of missing values must be imputed during the expectation (E) step. In this paper, we propose a new algorithm that is based on the EM algorithm but that efficiently computes the FIML estimates.  We include common factors in the complete data, as is the case with the ordinary EM algorithm, but we do not include the missing values on observed variables in the complete data.  Because of this, there is no need to impute the missing values in the E step. The proposed algorithm is applied to a real data set collected from a Web questionnaire that asks about first impressions of human; almost 90\% of the data values are missing.  
 Although the ordinary EM algorithm takes hours to run, our algorithm provides precise estimates in several tens of seconds.   

The second issue is the estimation accuracy of the FIML method: with the rate of missing values as large as 90\%, it is not clear whether the FIML procedure can yield a good estimator even if the number of observations is large, such as $N=2000$.  Although several researchers have discussed the effectiveness of the FIML estimator from both theoretical and numerical points of view (e.g., \citealp{finkbeiner1979estimation,lee1986estimation,enders2001relative,enders2001primer}), the rates of missing values they considered were not very large (typically, about 30\%).   In order to investigate how well the FIML method performs when the majority of data values are missing, we conducted Monte Carlo simulations under a wide variety of sample sizes.
 
The remainder of this paper is organized as follows: Section 2 defines the factor analysis model and notation, and briefly describes the FIML estimation procedure. In Section 3, we present the ordinary EM algorithms for FIML estimation, and we then modify the algorithm to improve the computational speed.  Section 4 presents an application of the proposed algorithm to data from a Web-based questionnaire.  In Section 5, a Monte Carlo simulation is conducted to investigate the effectiveness of the FIML procedure.  Some concluding remarks are given in Section 6. 

\section{FIML estimation in factor analysis}
Let $\bm{X}=(X_1,\dots,X_p)^T$ be a $p$-dimensional random vector with mean vector $\bm{\mu} $ and variance-covariance matrix $\bm{\Sigma}$. The factor analysis model (e.g., \citealp{mulaik2010foundations}) is
\begin{equation*}
\bm{X} =\bm{\mu} + \bm{\Lambda} \bm{F}+\bm{\varepsilon}  , \label{model1}
\end{equation*}
where $\bm{\Lambda} =(\lambda_{ij})$ is a $p \times m$  matrix of factor loadings, and $\bm{F} = (F_1,\cdots,F_m)^T$ and $\bm{\varepsilon}  = (\varepsilon_1,\cdots, \varepsilon_p)^T$ are unobservable random vectors. The elements of $\bm{F}$  and $\bm{\varepsilon}$  are called common factors and unique factors, respectively. It is assumed that the common factors $\bm{F}$ and the unique factors $\bm{\varepsilon}$ are multivariate-normally distributed with  $\E(\bm{F} )=\bm{0}$,  $\E(\bm{\varepsilon} )=\mathbf{0}$,  $\E(\bm{F}\bm{F}^T)=\bm{I}_m$, $\E(\bm{\varepsilon} \bm{\varepsilon} ^T)=\bm{\Psi} $, and are independent (i.e., $\E(\bm{F} \bm{\varepsilon} ^T)=\bm{O}$), where $\bm{I}_m$ is the identity matrix of order $m$, and $\bm{\Psi} $ is a $p \times p$ diagonal matrix in which the $i$-th diagonal element is $\psi_i$, which is called a unique variance.  Under these assumptions, the random vector $\bm{X}$ is multivariate-normally distributed with mean vector $\bm{\mu}$, and variance-covariance matrix $\bm{\Sigma} = \bm{\Lambda} \bm{\Lambda}^T+\bm{\Psi}$.   Note that the factor loadings have a rotational indeterminacy, because both $\bm{\Lambda} $ and  $\bm{\Lambda} \mathbf{T}$ generate the same covariance matrix $\mathbf{\Sigma}$, where $\mathbf{T}$ is an arbitrary orthogonal matrix.

We consider the case where the data values are partially observed.   Let $\bm{x}_1,\cdots,\bm{x}_N$ be $N$ sets of ``complete" data  drawn from $N_p(\bm{\mu},\bm{\Sigma})$ with $\bm{\Sigma} = \bm{\Lambda} \bm{\Lambda}^T+\bm{\Psi}$, which would occur in the absence of missing values.  The complete data vector $\bm{x}_n$ can be expressed as $\bm{x}_n=(\bm{x}_{[n]},\bm{x}_{-[n]})$, where $\bm{x}_{[n]}$ (or $\bm{x}_{-[n]}$) are observable (missing) values for case $n$.  Let $\bm{\mu}_{[n]}, \bm{\Lambda}_{[n]}$,  and $\bm{\Psi}_{[n]}$ denote model parameters based only on variables that are observed for case $n$.    The mean vector and covariance matrix based on the observed values $\bm{x}_{[n]}$ can be written as $\bm{\mu}_{[n]}$ and $\mathbf{\Sigma}_{[n]}=\bm{\Lambda}_{[n]} \bm{\Lambda}_{[n]}^T+\bm{\Psi}_{[n]}$, respectively.  The full information log likelihood function is then given by
\begin{equation}
\ell(\bm{\mu},\bm{\Lambda},\bm{\Psi})= -\frac{1}{2}\sum_{n=1}^N \bigg\{ p\log(2\pi)+\log |\mathbf{\Sigma}_{[n]}| + (\bm{x}_{[n]}-\bm{\mu}_{[n]})^T\mathbf{\Sigma}_{[n]}^{-1} (\bm{x}_{[n]}-\bm{\mu}_{[n]}) \bigg\} .\label{taisuuyuudo}
\end{equation}
The FIML estimates of $\bm{\mu}$, ${\bm{\Lambda} }$, and ${\bm{\Psi} }$ are given as the solutions of $\partial \ell / \partial \bm{\mu}  = \mathbf{0}$,  $\partial \ell / \partial \bm{\Lambda}  = \bm{O}$, and $\partial \ell / \partial \bm{\Psi}  = \bm{O}$, respectively. Since the solutions cannot be expressed in a closed form, we need to use an iterative algorithm, such as a quasi-Newton method or an EM algorithm.

\section{EM algorithms for FIML estimation}
In this section, we describe the ordinary EM algorithm for FIML estimations (e.g., \citealp{little1987statistical,jamshidian1997algorithm,liu1998maximum}); in this approach, both common factors and missing values on observed variables are included in the complete data.  In practical situations, however, the ordinary EM algorithm can be slow when the number of missing values on observed variables is large.  In order to handle this problem, we propose much more efficient algorithm.  A significant improvement in the computational speed is realized by not treating the missing values on observed variables as a part of complete data.   We call this approach the modified EM algorithm, and the details of the algorithm are given in Section \ref{sec:modified EM}. In Section \ref{sec:Comparison of computational cost}, we then discuss the computational complexity of matrix operations in order to compare the computational loads of the ordinary and the modified EM algorithms.

\subsection{Ordinary EM algorithm} \label{sec:ordinary EM}
The complete data log likelihood function $\l_{\rho}^{C} (\bm{\mu},\bm{\Lambda},\bm{\Psi})$ is expressed as
\begin{equation*}
\l_{\rho}^{C} (\bm{\mu},\bm{\Lambda},\bm{\Psi}) = \sum_{n=1}^N \log f(\bm{x}_n,\bm{f}_n),
\end{equation*}
where the density function $f(\bm{x}_n,\bm{f}_n)$ is defined by
\begin{equation*}
f(\bm{x}_n,\bm{f}_n) = \prod_{i=1}^p \left[  (2\pi\psi_i)^{-1/2} \exp \left\{ - \frac{ (x_{ni}-\mu_{ni} - \bm{\lambda}_i^T\bm{f}_n )^2}{2\psi_i}  \right\} \right]  (2\pi)^{-m/2}\exp \left( - \frac{\| \bm{f}_n \|^2}{2} \right).
\end{equation*}
Then, we have
\begin{eqnarray*}
\l_{\rho}^{C} &=&- \frac{N}{2} \sum_{i=1}^p \log \psi_i- \frac{1}{2} {\rm tr}\left\{ \bm{\Psi}^{-1} \sum_{n=1}^N(\bm{x}_n - \bm{\mu}-\bm{\Lambda}\bm{f}_n)(\bm{x}_n - \bm{\mu}-\bm{\Lambda}\bm{f}_n)^T \right\}  + C  \\
&=& - \frac{N}{2} \sum_{i=1}^p \log \psi_i - \frac{1}{2} {\rm tr}\left[ \bm{\Psi}^{-1} \sum_{n=1}^N 
\left\{ \bm{x}_n - ( \bm{\mu},\bm{\Lambda})
\begin{pmatrix}
 1\\
\bm{f}_n
 \end{pmatrix}
\right\}
\left\{ \bm{x}_n - ( \bm{\mu},\bm{\Lambda})
\begin{pmatrix}
 1\\
\bm{f}_n
 \end{pmatrix}
\right\}^T
\right] +C \\
&=& - \frac{N}{2} \sum_{i=1}^p \log \psi_i - \frac{1}{2} {\rm tr}\left[ \bm{\Psi}^{-1} \left\{ \bm{S}_{\bm{x}\bm{x}} -2  ( \bm{\mu},\bm{\Lambda}) \bm{S}_{\bm{f}^*\bm{x}} +( \bm{\mu},\bm{\Lambda})   \bm{S}_{\bm{f}^*\bm{f}^*}   
\begin{pmatrix}
 \bm{\mu}^T\\
\bm{\Lambda}^T
 \end{pmatrix}
\right\}
\right]  +C,
\end{eqnarray*}
where $C$ is a constant value and
$$
\bm{S}_{\bm{x}\bm{x}} = \sum_{n=1}^N\bm{x}_n\bm{x}_n^T, \quad 
\bm{S}_{\bm{f}^*\bm{x}} = \sum_{n=1}^N
\begin{pmatrix}
 1\\
\bm{f}_n
 \end{pmatrix}
\bm{x}_n^T, \quad 
\bm{S}_{\bm{f}^*\bm{f}^*} = \sum_{n=1}^N
\begin{pmatrix}
 1\\
\bm{f}_n
 \end{pmatrix}
(1,\bm{f}_n^T).
$$

\subsubsection*{E step} 
We compute the expectation of the sufficient statistics $\hat{\bm{S}}_{\bm{x}\bm{x}}=E[\bm{S}_{\bm{x}\bm{x}}|\bm{x}_{[1]},\dots, \bm{x}_{[N]},\hat{\bm{\theta}}]$, $\hat{\bm{S}}_{\bm{f}^*\bm{x}}=E[\bm{S}_{\bm{f}^*\bm{x}}|\bm{x}_{[1]},\dots, \bm{x}_{[N]},\hat{\bm{\theta}}]$, $\hat{\bm{S}}_{\bm{f}^*\bm{f}^*}=E[\bm{S}_{\bm{f}^*\bm{f}^*}|\bm{x}_{[1]},\dots, \bm{x}_{[N]},\hat{\bm{\theta}}]$ from the joint distribution of $(\bm{x}_n,\bm{f}_n)$ given $\bm{\theta}$:
\begin{equation}
\begin{pmatrix}
 \bm{x}_n\\
\bm{f}_n
 \end{pmatrix}
 \bigg|\bm{\theta}
 \sim
 N_{p+m}\left( 
 \begin{pmatrix}
\bm{\mu}\\
\bm{0}
 \end{pmatrix} , 
 \begin{bmatrix}
 \bm{\Lambda} \bm{\Lambda}^T + \bm{\Psi} & \bm{\Lambda}\\
 \bm{\Lambda}^T & \bm{I}
 \end{bmatrix}
  \right).\label{xn_fn_distribution}
\end{equation}
The joint distribution of $(\bm{x}_n,\bm{f}_n)$, given the observed values $\bm{x}_{[n]}$, can be obtained by using the standard methodology of a conditional Gaussian distribution.  Let $\bm{z}_n = (\bm{x}_{-[n]}^T,\bm{f}_n^T)^T$.   We set
\begin{equation*}
\begin{pmatrix}
\bm{z}_n\\
\bm{x}_{[n]}
\end{pmatrix}
\sim N \left(
\begin{pmatrix}
\bm{\mu}_{-[n]}\\
\bm{\mu}_{[n]}
\end{pmatrix},
\begin{pmatrix}
\bm{\Sigma}_{-[n],-[n]}&\bm{\Sigma}_{-[n],[n]}\\
\bm{\Sigma}_{[n],-[n]}&\bm{\Sigma}_{[n],[n]}\\
\end{pmatrix}
\right) .
\end{equation*}
The conditional distribution of $\bm{z}_n$ is given by
\begin{eqnarray*}
\bm{z}_n|\bm{x}_{[n]} &\sim& N(\bm{\mu}_{-[n]|[n]},\bm{\Omega}_{-[n],-[n]}^{-1}),\\
\bm{\mu}_{-[n]|[n]} &=& \bm{\mu}_{-[n]} - \bm{\Omega}_{-[n],-[n]}^{-1}\bm{\Omega}_{-[n],[n]}(\bm{x}_{[n]} - \bm{\mu}_{[n]}),\\
\begin{pmatrix}
\bm{\Omega}_{-[n],-[n]}&\bm{\Omega}_{-[n],[n]}\\
\bm{\Omega}_{[n],-[n]}&\bm{\Omega}_{[n],[n]}\\
\end{pmatrix}
&=&
\begin{pmatrix}
\bm{\Sigma}_{-[n],-[n]}&\bm{\Sigma}_{-[n],[n]}\\
\bm{\Sigma}_{[n],-[n]}&\bm{\Sigma}_{[n],[n]}\\
\end{pmatrix}^{-1}.
\end{eqnarray*}
On the other hand, $\bm{x}_{[n]}|\bm{x}_{[n]} \sim N(\bm{x}_{[n]} ,\bm{O}) $. Then, we can compute the conditional distribution
\begin{eqnarray}
\begin{pmatrix}
\bm{x}_n\\
\bm{f}_n
\end{pmatrix}
\bigg| \bm{x}_{[n]} \sim N\left(
\begin{pmatrix}
\hat{\bm{x}}_n\\
\hat{\bm{f}}_n\\
\end{pmatrix},
\begin{pmatrix}
\hat{\bm{V}}_{\bm{x_n}\bm{x}_n^T}&\hat{\bm{V}}_{\bm{x_n}\bm{f}_n^T}\\
\hat{\bm{V}}_{\bm{f_n}\bm{x}_n^T}&\hat{\bm{V}}_{\bm{f_n}\bm{f}_n^T}\\
\end{pmatrix}
\right).\label{posteriorall}
\end{eqnarray}
The sufficient statistics $\hat{\bm{S}}_{\bm{x}\bm{x}}$, $\hat{\bm{S}}_{\bm{f}^*\bm{x}}$, and $\hat{\bm{S}}_{\bm{f}^*\bm{f}^*}$ are expressed as
\begin{eqnarray}
\hat{\bm{S}}_{\bm{x}\bm{x}}&=&\sum_{n=1}^N(\hat{\bm{x}}_n\hat{\bm{x}}_n^T + \hat{\bm{V}}_{\bm{x}_n\bm{x}_n^T}), \quad
\hat{\bm{S}}_{\bm{f}^*\bm{x}}=\sum_{n=1}^N
\begin{pmatrix}
\hat{\bm{x}}_n^T\\
\hat{\bm{f}}_n\hat{\bm{x}}_n^T + \hat{\bm{V}}_{\bm{f}_n\bm{x}_n^T}
\end{pmatrix}, \cr 
\hat{\bm{S}}_{\bm{f}^*\bm{f}}&=&\sum_{n=1}^N
\begin{pmatrix}
1 & \hat{\bm{f}}_n^T\\
\hat{\bm{f}}_n & \widehat{\bm{f}_n \bm{f}_n^T}
\end{pmatrix},
\nonumber
\end{eqnarray}
where $\widehat{\bm{f}_n \bm{f}_n^T} = \hat{\bm{f}}_n\hat{\bm{f}}_n^T +  \hat{\bm{V}}_{\bm{f}_n\bm{f}_n^T}$.
\subsubsection*{M step}
In the maximization (M) step, we maximize the complete data log likelihood function.  By taking the derivative with respect to $(\bm{\mu},\bm{\Lambda})$ and $\bm{\Psi}$, we have
\begin{eqnarray*}
\frac{\partial E[l_{\rho}]}{\partial (\bm{\mu},\bm{\Lambda})} &=& -\frac{1}{2} (-2\bm{\Psi}^{-1}\hat{\bm{S}}_{\bm{f}^*\bm{x}}^T + 2 \bm{\Psi}^{-1}(\bm{\mu},\bm{\Lambda})\hat{\bm{S}}_{\bm{f}^*\bm{f}^*}^T),\\
\frac{\partial E[l_{\rho}]}{\partial \bm{\Psi}^{-1}} &=& \frac{N}{2} {\rm diag}(\bm{\Psi})  - \frac{1}{2} {\rm diag} \left[\hat{\bm{S}}_{\bm{x}\bm{x}} -2  ( \bm{\mu},\bm{\Lambda}) \hat{\bm{S}}_{\bm{f}^*\bm{x}} +( \bm{\mu},\bm{\Lambda})   \hat{\bm{S}}_{\bm{f}^*\bm{f}^*}   
\begin{pmatrix}
 \bm{\mu}^T\\
\bm{\Lambda}^T
 \end{pmatrix}
\right].
\end{eqnarray*}
The solution is given by
\begin{eqnarray*}
 (\bm{\mu},\bm{\Lambda}) &=& \hat{\bm{S}}_{\bm{f}^*\bm{x}}^T \hat{\bm{S}}_{\bm{f}^*\bm{f}^*}^{-1}, \label{mstep1-1}\\
\bm{\Psi} &=& \frac{1}{N} {\rm diag} \left[ \hat{\bm{S}}_{\bm{x}\bm{x}} -2  ( \bm{\mu},\bm{\Lambda}) \hat{\bm{S}}_{\bm{f}^*\bm{x}} +( \bm{\mu},\bm{\Lambda})   \hat{\bm{S}}_{\bm{f}^*\bm{f}^*}   
\begin{pmatrix}
 \bm{\mu}^T\\
\bm{\Lambda}^T
 \end{pmatrix}
\right].\label{mstep1-2}
\end{eqnarray*}

\subsection{Modified EM algorithm}\label{sec:modified EM}
When the number of missing values is very large, the ordinary EM algorithm in Section \ref{sec:ordinary EM} becomes inefficient, because we must impute a number of missing values in the E step.  In order to overcome this problem, we introduce a modified algorithm. An important point in our algorithm is that the missing values on observed variable $\bm{x}_{-[n]}$  are {\it not} included in the complete data. In this case, the complete data log likelihood function is given by
\begin{eqnarray*}
\l_{\rho}^{C^*}  &=& -\frac{1}{2}  \sum_{n \in n{\rm obs}(i)}  \sum_{i=1}^p\log \psi_i \\
&&- \frac{1}{2}   \sum_{n \in n{\rm obs}(i)}  \sum_{i=1}^p \frac{(x_{ni}-\mu_{i})^2 - 2(x_{ni}-\mu_i)\bm{\lambda}_i^T\bm{f}_n+ \bm{\lambda}_i^T \bm{f}_n\bm{f}_n^T\bm{\lambda}_i}{\psi_i}, 
\end{eqnarray*}
where $n{\rm obs}(i) =\{n \in \{1,\dots, N\} \mid \mbox{$i$-th variable is observed.}\}$
\subsubsection*{E step}
We need to compute the expected values of only the common factors given the observed data, i.e., $\hat{\bm{f}}_n$ and  $\hat{\bm{V}}_{\bm{f}_n\bm{f}_n^T}$  in (\ref{posteriorall}).
\subsubsection*{M step}
We can take the derivatives with respect to $\bm{\mu}$, $\bm{\Lambda}$, and $\bm{\Psi}$, which are written as 
\begin{eqnarray*}
\frac{\partial E[l_{\rho}^{C^*}]}{\partial \mu_i} &=& -\frac{1}{2 \psi_i} \sum_{n \in n{\rm obs}(i)} \left\{ -2(x_{ni}-\mu_i) +2 \bm{\lambda}_i^T\hat{\bm{f}_n}) \right\},\\
\frac{\partial E[l_{\rho}^{C^*}]}{\partial \bm{\lambda}_i} &=& -\frac{1}{2 \psi_i} \sum_{n \in n{\rm obs}(i)} \left\{ -2(x_{ni}-\mu_i) \hat{\bm{f}_n} + 2  \widehat{\bm{f}_n \bm{f}_n^T}\bm{\lambda}_i ) \right\},\\
\frac{\partial E[l_{\rho}^{C^*}]}{\partial \psi_i^{-1}} &=& \frac{\# n{\rm obs}(i)}{2} \psi_i - \frac{1}{2} \sum_{n \in n{\rm obs}(i)} \left\{ (x_{ni}-\mu_{i})^2 - 2(x_{ni}-\mu_i)\bm{\lambda}_i^T\hat{\bm{f}}_n+ \bm{\lambda}_i^T \widehat{\bm{f}_n \bm{f}_n^T}\bm{\lambda}_i \right\}.
\end{eqnarray*}
The solutions are
\begin{eqnarray*}
 \mu_i&=& \frac{1}{\# n{\rm obs}(i)} \sum_{n \in n{\rm obs}(i)} (x_{ni} - \bm{\lambda}_i^T\hat{\bm{f}_n} ),\label{mstep2-1}\\
\bm{\lambda}_i &=& \left\{ \sum_{n \in n{\rm obs}(i)} \widehat{\bm{f}_n \bm{f}_n^T} \right\}^{-1} \left\{ \sum_{n \in n{\rm obs}(i)}  (x_{ni}-\mu_i) \hat{\bm{f}_n}  \right\},  \label{mstep2-2} \\
{\psi}_i &=& \frac{1}{\# n{\rm obs}(i)} \sum_{n \in n{\rm obs}(i)}\left\{ (x_{ni}-\mu_{i})^2 - 2(x_{ni}-\mu_i)\bm{\lambda}_i^T\hat{\bm{f}}_n+ \bm{\lambda}_i^T \widehat{\bm{f}_n \bm{f}_n^T}\bm{\lambda}_i \right\}. \label{mstep2-3}
\end{eqnarray*}
\subsection{Computational complexity of matrix operations}\label{sec:Comparison of computational cost}
In this section, we discuss the computational complexity of the matrix operations for each algorithm.  For ease of comprehension, we assume that the number of missing (or observed) variables, say, $p_{\rm mis}$ (or $p_{\rm obs}$), is constant for all observations. Note that the computational complexity independent of this assumption can be discussed in the same manner. 

Assume that a massive amount of data is missing, i.e., $p_{\rm mis} \approx p$, and $m$ is sufficiently small.   In the E step of the ordinary EM algorithm, the operation $\bm{\Omega}_{-[n],-[n]}^{-1}$ is almost $O(p^2)$.  To show this, first, we calculate the inverse of the covariance matrix of the joint distribution $(\bm{x}_n^T,\bm{f}_{n}^T)^T$ in (\ref{xn_fn_distribution})
\begin{equation*}
   \begin{bmatrix}
 \bm{\Lambda} \bm{\Lambda}^T + \bm{\Psi} & \bm{\Lambda}\\
 \bm{\Lambda}^T & \bm{I}
 \end{bmatrix}^{-1} 
 = 
   \begin{bmatrix}
 \bm{\Psi}^{-1} & -\bm{\Psi}^{-1} \bm{\Lambda}\\
 -\bm{\Lambda}^T\bm{\Psi}^{-1} & \bm{M}
 \end{bmatrix},
\end{equation*}
where $\bm{M} = \bm{\Lambda}^T\bm{\Psi}^{-1}\bm{\Lambda} + \bm{I}$.  Thus, we have 
\begin{equation*}
  \bm{\Omega}_{-[n],-[n]}^{-1} = \bm{\Lambda}_{-[n]} (\bm{M} - \bm{\Lambda}_{-[n]}^{T} \bm{\Psi}_{-[n]}^{-1}\bm{\Lambda}_{-[n]})^{-1} \bm{\Lambda}_{-[n]}^{T},
\end{equation*}
which requires $O(p^2)$ when $m$ is small.  The computational complexity of the E step is then given by  $O(Np^2)$.  The computational complexity of the M step is  $O(p)$, which is sufficiently small compared with that of the E step.

On the other hand, the modified EM algorithm is much more efficient: the operation needs only $O(Np_{\rm obs}^2)$.  Furthermore, with a large rate of missing values, we found that the number of iterations in the modified EM algorithm tends to be much smaller than that of the ordinary EM algorithm, as shown in the simulation study in Section \ref{sec.timing}.  However, we do not yet have mathematical support for this claim.  We would like to consider this as a future research topic.

\section{Analysis of data from a Web-based questionnaire on first impressions} \label{sec:realdata}
We now explore the underlying factor structure of personal assessments of first impressions, based on data from a Web-based questionnaire.  The responders were asked to evaluate four virtual people based on several paired adjectives (e.g., pleasant - unpleasant) on a scale of 1 to 5.  In order to use a wide variety of personal assessment measures for investigating the underlying structure of first impressions, we prepared 94 measures.  Answering 94 items is a heavy load, so the following procedure was carried out:
\begin{enumerate}
  \item Before the four virtual people were displayed, the participants selected four assessment measures (selective measures) that they used in their daily life.
  \item The participants evaluated the four virtual people based on the four selective measures and an additional six assessment measures that were assigned to all participants (common measures).   The six common measure  are as follows: ``pleasant - unpleasant", ``friendly - unfriendly", ``careful - hasty", ``sensible - insensible", ``active - passive", and ``confident - unconfident."
\end{enumerate}
Each participant only selected $10$ $(=4+6)$ items out of 94 items, so that almost 90\% of the data values were missing.  Because 8544 participants appropriately completed the questionnaire for four virtual people, the number of observations is $8544 \times 4=34176$.  The number of factors was set to be $m=3$, because \citet{rosenberg1968multidimensional} described personality impressions as being based on a three-dimensional configuration.  
  
First, the computational time based on the ordinary EM algorithm described in Section \ref{sec:ordinary EM} was compared with that of the modified EM algorithm described in Section \ref{sec:modified EM}. A quasi-Newton method was also compared; the inverse of the Hessian matrix was approximated by the Broyden--Fletcher--Goldfarb--Shanno (BFGS) algorithm.  The quasi-Newton method uses the full information likelihood function in (\ref{taisuuyuudo}) and its first derivatives given by
\begin{eqnarray*}
\dfrac{\partial \ell(\bm{\mu},\bm{\Lambda},\bm{\Psi})}{\partial \bm{\mu}}&=& \sum_{n=1}^N \mathbf{\Sigma}_{[n]}^{-1}(\bm{x}_{[n]} - \bm{\mu}_{[n]}),  \vspace{0.5em} \label{mubibun=02-1-2} \\
\dfrac{\partial \ell(\bm{\mu},\bm{\Lambda},\bm{\Psi})}{\partial \bm{\Lambda}}&=&  \sum_{n=1}^N(\mathbf{\Sigma}_{[n]}^{-1}\bm{x}_{[n]}\bm{x}_{[n]}^T\mathbf{\Sigma}_{[n]}^{-1} - \mathbf{\Sigma}_{[n]}^{-1})\bm{\Lambda}_{[n]},  \vspace{0.5em} \label{Lbibun=02-1-2} \\
\dfrac{\partial \ell(\bm{\mu},\bm{\Lambda},\bm{\Psi})}{\partial \bm{\Psi}} &=& \frac{1}{2} \sum_{n=1}^N \mathrm{diag}(\mathbf{\Sigma}_{[n]}^{-1}\bm{x}_{[n]}\bm{x}_{[n]}^T\mathbf{\Sigma}_{[n]}^{-1} - \mathbf{\Sigma}_{[n]}^{-1}). \label{Psibibun=02-1-2}
\end{eqnarray*}
Note that the quasi-Newton algorithm can be inefficient when the number of observations is very large, because the covariance matrix $\mathbf{\Sigma}_{[n]}$ and its inverse must be computed for each case $n$.

We computed the average time for 10 runs using different initial values. All computations were carried out with Windows 8 and an Intel Core i7 3.4 GH processor. The program was written in {\tt R} using {\tt C}. For the quasi-Newton method via BFGS optimization, we used the {\tt vmmin} function called by the {\tt optim} function in {\tt R}.  The result was:
\begin{itemize}
\item modified EM algorithm: \quad 7.81 seconds,
\item ordinary EM algorithm: \quad 9.00 hours,
\item quasi-Newton method: \quad \ \  25.8 minutes.
\end{itemize}
Our algorithm was considerably faster than the two existing methods.  Note that the EM algorithm converged to the FIML estimates for all 10 initial values, whereas the quasi-Newton algorithm diverged for 2 out of 10 initial values.  Thus, the quasi-Newton algorithm may be unstable compared with the EM algorithm.

Next, the loading matrix was rotated by the promax method \citep{hendrickson1964promax} to interpret the estimated common factors.  The estimated factor loadings and unique variances are shown in Appendix A.  The results show that the FIML procedure was able to produce the following three interpretable common factors:  {\it personality}, {\it intelligence}, and {\it activeness}.  

Although the estimated model is interpretable, it is not clear yet whether the FIML can achieve good estimation accuracy when the missing value rate is as large as 90\% even if the number of observations is as large as $N=30000$.  In order to investigate how well the FIML method performs when the majority of the data values are missing, we conduct Monte Carlo simulations under a wide variety of sample sizes, as shown in the next section.

\section{Monte Carlo Simulations}\label{sec:simulation}
In the simulations, we used the following loading matrix and unique variances:
\begin{eqnarray*}
\bm{\Lambda} &=&  (\underbrace{0.8 \bm{I}_3, \ 0.8 \bm{I}_3, \ \dots, \ 0.8 \bm{I}_3}_{30})^T, \quad  \bm{\Psi} = {\rm diag}(\bm{I} - \bm{\Lambda} \bm{\Lambda}^T ).
\end{eqnarray*}
In this case, $p=90$ and $m=3$.  The model was estimated by the maximum likelihood method under the rotational restriction that the upper triangular matrix of the loading matrix is zero, i.e., $\lambda_{ij}=0 $ ($j>i$) (e.g., \citealp{anderson1956statistical}).  The aim of this simulation study is to  (i) investigate how well the FIML method performs when the majority of the data are missing, and (ii) compare the computation times of the quasi-Newton method, the ordinary EM algorithm, and the modified EM algorithm.

\subsection{Investigation of performance of the FIML estimation}\label{sec:simulation}
First, we investigated the performance of the FIML procedure when a large number of data values were missing.  The number of observations was the sequence of twenty integers decreasing on the log scale from $N=40000$  to  $N=200$.    We first generated the common factors and unique factors by using $\bm{f}_n \sim N(\bm{0}, \bm{I}_3)$ and $\bm{\varepsilon}_n \sim N(\bm{0},\bm{\Psi})$, and then the complete data was created by $\bm{x}_n = \bm{\Lambda} \bm{f}_n  + \bm{\varepsilon}_n $ $(n=1,\dots,N)$. 

At each observation, we chose (approximately) $q$ variables and eliminated them.   The mechanism for choosing which values to eliminate was assumed to be either missing completely at random (MCAR) or not missing at random (NMAR), as follows: 
\begin{description}
\item[MCAR:] We randomly chose $q$ variables and set these as the missing values.  
\item[NMAR:] For the $i$-th variable of the $n$-th subject, we calculated the value based on the logistic function $p_{in} = 1/(1+\exp(-\alpha \bm{\lambda}_i^T\bm{f}_n))$, and then  the missing indicator values for $x_{in}$ were generated from the Bernoulli distribution with probability $p_{in}$.  The value of $\alpha$ was chosen so that the mean value of the $p_{in}$ approximates the missing rate, i.e., $\sum_{i,n}p_{in}/(Np) \approx q/p$.
\end{description}
Note that the MCAR assumption is a special case of MAR, so the FIML procedure produces a consistent estimator under the assumption of MCAR.  On the other hand, the NMAR assumption leads to an inconsistent estimator.  

In each case, the first six items were assumed to be ``common measures" that were not allowed to be missing (i.e., all subjects must answer these six questions).  To investigate the effectiveness of the common measures, we also estimated the model without the common measures, i.e., the common measures were eliminated and the model was estimated by using 84  ($=90-6$) variables.  This procedure was repeated 1000 times. Figure \ref{fig:simulation_msebias} shows the square root of the mean squared error (${\rm sqrtMSE}$) and the bias (${\rm sqrtBIAS}$) of the estimator $\bm{\Lambda}$ defined by
\begin{eqnarray*}
{\rm sqrtMSE} &=&\sqrt{ \frac{1}{1000r}  \sum_{j=1}^{\max(i,m)} \sum_{i=7}^p \sum_{s=1}^{1000} ( \hat{\lambda}_{ij}{(s)} - \lambda_{ij} )^2 }, \\ 
{\rm sqrtBIAS}& =&\sqrt{ \frac{1}{r}  \sum_{j=1}^{\max(i,m)} \sum_{i=7}^p  ( \bar{\lambda}_{ij} - \lambda_{ij})^2 },
\end{eqnarray*}
where $\hat{\lambda}_{ij}{(s)}$ is the maximum likelihood estimate for the $s$-th dataset, $\bar{\lambda}_{ij} = \sum_{s}\hat{\lambda}_{ij}{(s)}/1000$, and $r$ is the number of parameters of the last 84 rows of $\bm{\Lambda}$ given by $r=(p-6)m-m(m-1)/2$. Note that the first six rows of the loading matrix were not used to compute the sqrtBIAS and sqrtMSE; this is because we would like to investigate whether the common measures yield a good estimate of the parameters that correspond to the other 84 variables.  
\begin{figure*}[!t]
\begin{center}
    \includegraphics[width=135mm]{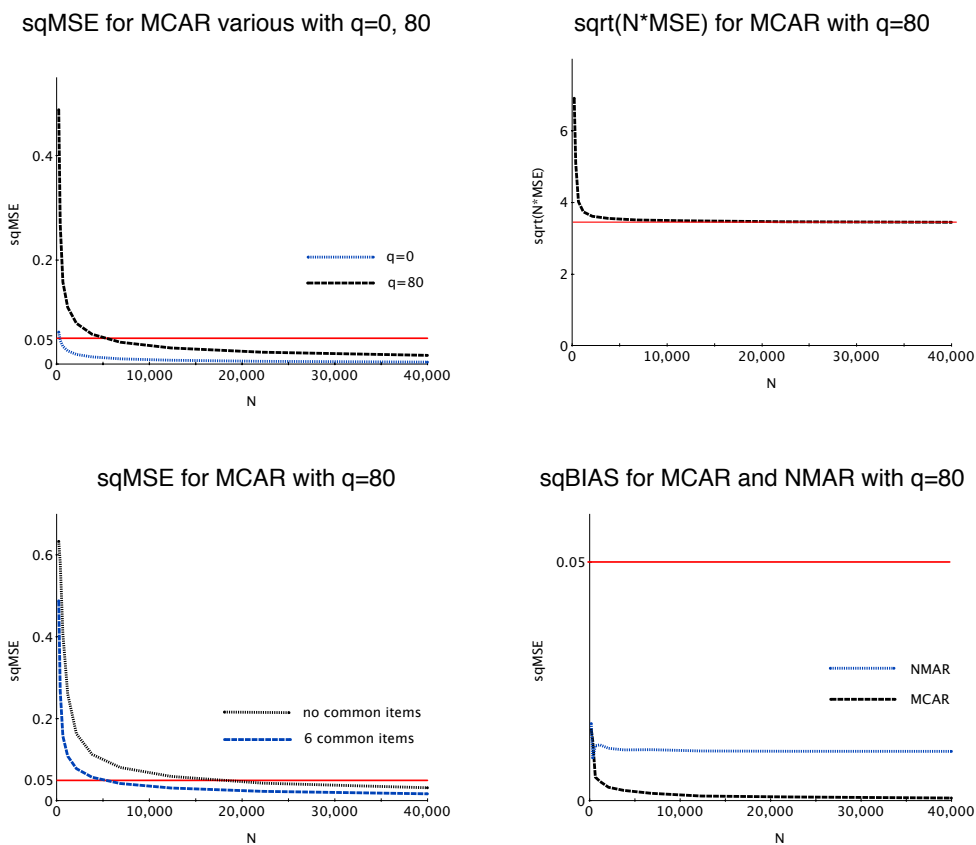}\vspace{5mm}
 \caption{Square root of the mean squared error (${\rm sqrtMSE}$) and bias (${\rm sqrtBIAS}$) of the estimator $\bm{\Lambda}$. The horizontal line shows ${\rm sqrtMSE}=0.05$, which might be small enough that the estimated model can be interpreted.  }
\label{fig:simulation_msebias}
\end{center}
\end{figure*}
The range of index $j$ is $\max(i,m)$ because the upper triangular matrix of the loading matrix is zero.  

We provide a detailed description and discussion of Figure \ref{fig:simulation_msebias}:
\begin{itemize}
\item The upper left panel shows the sqrtMSE for MCAR with $q=0$ and $q=80$. When $q = 80$, the missing value rate was about 90\%, which is similar to the setting used for the Web-based questionnaire data analysis described in Section \ref{sec:realdata}. The horizontal line shows ${\rm sqrtMSE}=0.05$, which may be small enough to correctly interpret the estimated model if the observed variables are scaled to have unit variance.  The sqrtMSE for $q=80$ was much larger than that for $q=0$ when $N<10000$.  We may need a large number of observations, such as $N=10000$, to obtain an accurate estimate when a massive amount of data is missing. 
\item The upper right panel depicts $\sqrt{N \cdot {\rm MSE}}$. It is well known that $\sqrt{{\rm MSE}}$ possesses  $\sqrt{N}$-consistency, so that $\sqrt{N \cdot {\rm MSE}}$ may be constant for large values of $N$.  We can see that the estimated MSE may be close to the true MSE when $N>20000$. When $N=20000$, the sqrtMSE was approximately $0.02$, which is small enough to correctly interpret the estimated model.   As a result, we may need $N>20000$ to produce an accurate estimation.
\item The lower left panel shows the sqrtMSE with six common measures and with no common measures.  This shows that the common measures play an important role in making the value of the sqrtMSE smaller.  
\item The lower right panel shows the sqrtBIAS for MCAR and NMAR.  This was done to investigate how well the FIML performs when the true missing mechanism is NMAR. When the missing mechanism is MCAR, the sqrtBIAS converges to zero, which means the FIML produces a consistent estimator.  On the other hand, the FIML estimates in the NMAR case are biased, so that the sqrtBIAS seems to converge to some small positive value when $N \rightarrow \infty$. However, the sqrtBIAS was approximately $0.01$, which may be sufficiently small compared with the sqrtMSE depicted in the left upper panel.   
\end{itemize}

We also computed the minimum number of observations required to satisfy ${\rm sqrtMSE}<0.05$, $0.025$ for various $q$; these are shown in Table \ref{table:sqrtMSE}. For example, when $q \ge 80$, we need at least $N = 20000$ observations to satisfy ${\rm sqrtMSE}<0.025$.

In the data from the Web-based questionnaire, as described in Section \ref{sec:realdata}, the number of observations was $N=34176$.  Therefore, the value of the ${\rm sqrtMSE}$ might be less than $0.025$, which is sufficiently small to correctly interpret the estimated model.     
\begin{table}[t]
\caption{Minimum number of observations that satisfy the sqrtMSE for various $q$. } \label{table:sqrtMSE}
\begin{center}
\begin{tabular}{rrrrrrrrrr}
  \hline
sqrtMSE & $q=0$ & $q=20$ & $q=40$ & $q=60$ & $q=70$ & $q=80$ \\ 
  \hline
0.025 & 1279  & 1460 & 1869 & 3134 & 5281 & 20056 \\ 
  0.05 & 321 & 385 & 516 & 605  & 1363 & 5329 \\ 
  \hline
\end{tabular}
\end{center}
\end{table}

\subsection{Comparison of computation times}\label{sec.timing}
We computed the computation time and the number of iterations for MCAR with common measures when $q=0,10,20,\dots,80$ and $N=2000$.  The other settings were the same as for the comparison of computation times in the analysis of actual data, as discussed in Section \ref{sec:realdata}.  Figure \ref{fig:simulation_timing} shows the computation times and the number of iterations, each averaged over 10 runs for each of the three algorithms (quasi-Newton method, ordinary EM, and modified EM).  Note that these algorithms converged to the same solutions when starting with the same initial values.   

From the results presented in Figure \ref{fig:simulation_timing}, we can see that
\begin{itemize}
\item  The modified EM algorithm was the fastest among the three algorithms when $q \ge 10$.  In particular, when $q=80$ (i.e., the majority of the data values were missing), the modified EM algorithm was 247 times faster than the ordinary EM algorithm, and 128 times faster than the quasi-Newton method.  
\item The number of iterations of the ordinary EM algorithm increased as the number of missing variables $q$ increased. This shows that the ordinary EM algorithm may be inefficient when the number of missing values is very large.  On the other hand, for both the quasi-Newton method and the modified EM algorithm, the number of iterations decreased as the number of missing values increased.  
\end{itemize}
\begin{figure*}[!t]
\begin{center}
    \includegraphics[width=135mm]{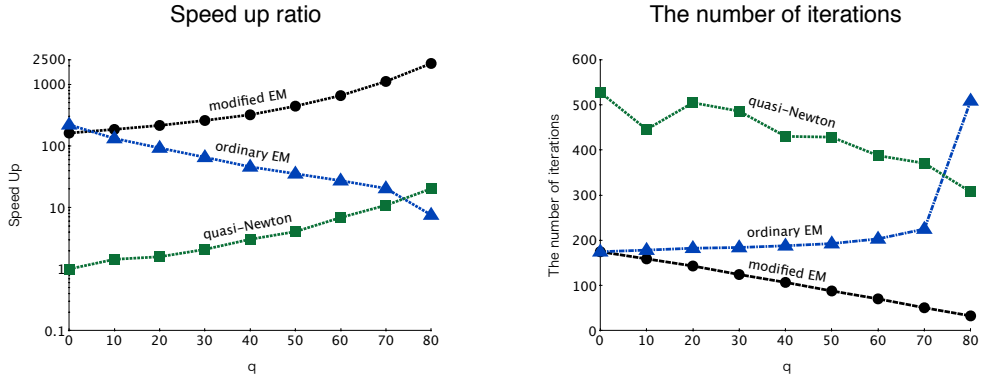}\hspace{5mm}
 \caption{Comparison of calculation time.  The left panel shows the speed up ratio; the baseline is the quasi-Newton method without missing data. The right panel depicts the number of iterations for each method.}
\label{fig:simulation_timing}
\end{center}
\end{figure*}

\section{Concluding remarks}
We presented a new FIML estimation algorithm that improves the computational speed of the ordinary EM algorithm. In the analysis of actual data, the proposed algorithm was considerably faster than the ordinary EM algorithm. We also conducted Monte Carlo simulations to investigate the performance of the FIML procedure. The results showed that several tens of thousands of observations may be necessary in order to obtain an accurate estimate when the rate of missing values was 90\%. 

Although the FIML procedure performed well even when the true missing data were NMAR based on the logistic function, various other NMAR cases were not explored (e.g., \citealp{yuan2009identifying,kano2011analysis}). As a future research topic, it would be interesting to explore the performance of FIML estimation and to determine algorithms that would be efficient for various NMAR cases.   Another topic would be to determine a much faster algorithm for high-dimensional sparse data, such as the Netflix Prize dataset \citep{bennett2007netflix}, which consists of $(N,p)=(480189,17700)$ with 99\% of the data missing.

 \appendix
\section{Estimates of factor loadings for the analysis of the Web-based questionnaire data}
The FIML estimates of the factor loadings and unique variances are shown in Tables \ref{tab:factor1},  \ref{tab:factor23}, and \ref{tab:unique}. The estimates of the factor loadings were rotated by the promax method \citep{hendrickson1964promax}. Table \ref{tab:factor1} shows the adjective pairs related to factor 1, and Table \ref{tab:factor23} presents the items related to factors 2 and 3. From 94 personality traits, the following three common factors were found: personality (Factor 1), intelligence (Factor 2), and activeness (Factor 3). Table \ref{tab:unique} shows the adjective pairs that possess large unique variances, which means that these items are not very closely related to these three factors.  
\begin{table}[!htdp]
\caption{Factor loadings for 28 items that possesses large absolute numbers for Factor 1.  The absolute values of factor loadings that are larger than 0.4 are in bold.}\label{tab:factor1}
\begin{center}
\begin{tabular}{lrrrr}
  \hline
    Adjective Pairs & Factor 1 & Factor 2 & Factor 3 & Uniquenesses \\
  \hline
    Pleasant $-$ Unpleasant & {\bf 0.706} & 0.154 & 0.061 & 0.262 \\
    Friendly $-$ Unfriendly & {\bf 0.791} & $-$0.022 & 0.150  & 0.266 \\
    Casual $-$ Formal & {\bf 0.691} & $-$0.221 & 0.172 & 0.575 \\
    Dishonest $-$ Honest & {\bf $-$0.479} & $-$0.333 & 0.238 & 0.536 \\
    Bad Feeling $-$ Good Feeling & {\bf $-$0.579} & $-$0.036 & $-$0.107 & 0.422 \\
    Obedient $-$ Disobedient & {\bf 0.729} & 0.101 & $-$0.160 & 0.459 \\
    Skeptical $-$ Credulous & {\bf $-$0.632} & 0.396 & $-$0.046 & 0.827 \\
    Honest $-$ Liar & {\bf 0.515} & 0.289 & $-$0.087 & 0.475 \\
    Modest $-$ Immodest & {\bf 0.527} & 0.323 & {\bf $-$0.462} & 0.507 \\
    Frank $-$ Formal & {\bf 0.624} & $-$0.169 & 0.345 & 0.383 \\
    Mild $-$ Intense & {\bf 0.681} & 0.240  & $-$0.396 & 0.420 \\
    Kind $-$ Unkind & {\bf 0.672} & 0.179 & $-$0.016 & 0.269 \\
    Sympathetic $-$ Unsympathetic & {\bf 0.683} & 0.250  & $-$0.143 & 0.377 \\
    Warm $-$ Cold & {\bf 0.810}  & 0.092 & $-$0.043 & 0.281 \\
    Acid $-$ Round & {\bf $-$0.643} & 0.012 & 0.305 & 0.559 \\
    Patient $-$ Impatient & {\bf 0.595} & 0.111 & $-$0.272 & 0.520 \\
    Soft $-$ Hard & {\bf 0.812} & $-$0.081 & $-$0.002 & 0.427 \\
    Tough $-$ Gentle & {\bf $-$0.546} & 0.222 & 0.384 & 0.615 \\
    Mean $-$ Nice & {\bf $-$0.547} & $-$0.052 & 0.195 & 0.454 \\
    Laid-back $-$ Rash & {\bf 0.723} & $-$0.058 & $-$0.339 & 0.529 \\
    Interesting $-$ Boring & {\bf 0.440}  & $-$0.113 & {\bf 0.438} & 0.515 \\
    Cheerful $-$ Depressing & {\bf 0.571} & $-$0.030 & 0.239 & 0.346 \\
    (Agree $-$ Disagree) with Each Other & {\bf 0.666} & 0.056 & 0.090  & 0.448 \\
    (Same $-$ Different) Ways of Thinking & {\bf 0.589} & 0.177 & 0.004 & 0.425 \\
    Empathetic $-$ Lack Empathy & {\bf 0.644} & 0.230  & 0.031 & 0.370 \\
    Feel at Ease $-$ Frustrating & {\bf 0.716} & 0.144 & $-$0.072 & 0.341 \\
    Safe $-$ Dangerous & {\bf 0.656} & 0.302 & $-$0.099 & 0.378 \\
    Friend $-$ Enemy & {\bf 0.656} & 0.114 & 0.029 & 0.317 \\
  \hline
\end{tabular}
\end{center}
\end{table}

\begin{table}[!htdp]
\caption{Factor loadings for 29 items that possesses large absolute numbers for Factors 2 and 3. The absolute values of factor loadings that are larger than 0.4 are in bold.}\label{tab:factor23}
\begin{center}
\begin{tabular}{lrrrr}
  \hline
    Adjective Pairs & Factor 1 & Factor 2 & Factor 3 & Uniquenesses \\
  \hline
    Careful $-$ Hasty & 0.193 & {\bf 0.531} & $-$0.169 & 0.345 \\
    Sensible $-$ Insensible & 0.313 & {\bf 0.516} & $-$0.017 & 0.295 \\
    Stable $-$ Unstable & 0.142 & {\bf 0.517} & 0.096 & 0.436 \\
    Neat $-$ Untidy & 0.195 & {\bf 0.606} & 0.041 & 0.400 \\
    Serious $-$ Frivolous & 0.398 & {\bf 0.518} & $-$0.128 & 0.341 \\
    Responsible $-$ Irresponsible & 0.171 & {\bf 0.621} & 0.028 & 0.417 \\
    Careful $-$ Careless & $-$0.060 & {\bf 0.667} & $-$0.060 & 0.391 \\
    Intellectual $-$ Sensuous  & 0.004 & {\bf 0.638} & 0.039 & 0.468 \\
    Mature $-$ Childish & 0.061 & {\bf 0.578} & 0.048 & 0.452 \\
    Calm $-$ Passionate & 0.081 & {\bf 0.549} & $-$0.227 & 0.531 \\
    Logical $-$ Emotional & $-$0.158 & {\bf 0.715} & $-$0.025 & 0.506 \\
    Respected $-$ Disrespectful  & 0.382 & {\bf 0.418} & 0.082 & 0.378 \\
    Active $-$ Passive & $-$0.166 & 0.103 & {\bf 0.779} & 0.231 \\
    Confident $-$ Unconfident & $-$0.220 & 0.208 & {\bf 0.725} & 0.256 \\
    Sober $-$ Flashy & 0.339 & 0.371 & {\bf $-$0.495} & 0.578 \\
    Healthy $-$ Sickly & 0.150  & $-$0.036 & {\bf 0.638} & 0.299 \\
    Strong $-$ Weak & $-$0.255 & 0.102 & {\bf 0.784} & 0.365 \\
    Reliable $-$ Unreliable & 0.020  & 0.299 & {\bf 0.548} & 0.413 \\
    Bold $-$ Timid & $-$0.019 & $-$0.061 & {\bf 0.731} & 0.481 \\
    Clear $-$ Vague & $-$0.193 & 0.236 & {\bf 0.660}  & 0.437 \\
    Loud $-$ Quiet & 0.105 & $-$0.140 & {\bf 0.724} & 0.425 \\
    Extrovert $-$ Introvert & $-$0.039 & $-$0.029 & {\bf 0.771} & 0.347 \\
    Talkative $-$ Taciturn & 0.084 & $-$0.161 & {\bf 0.712} & 0.480 \\
    Inner $-$ Outward & 0.270  & 0.318 & {\bf $-$0.552} & 0.606 \\
    Exhibitionist $-$ Quiet & $-$0.235 & $-$0.100  & {\bf 0.810}  & 0.476 \\
    Bright $-$ Dark & 0.382 & $-$0.120 & {\bf 0.553} & 0.347 \\
    Cheerful $-$ Dismal & 0.439 & $-$0.134 & {\bf 0.504} & 0.297 \\
    Rich $-$ Poor & 0.123 & 0.337 & 0.397 & 0.402 \\
    Superior $-$ Inferior & $-$0.094 & {\bf 0.422} & {\bf 0.452} & 0.342 \\
      \hline
\end{tabular}
\end{center}
\end{table}

\begin{table}[!htdp]
\caption{Factor loadings for 37 items that possesses large numbers of unique variances. The absolute values of factor loadings that are larger than 0.4 are in bold.}\label{tab:unique}
\begin{center}
\begin{tabular}{lrrrr}
  \hline
    Adjective Pairs & Factor 1 & Factor 2 & Factor 3 & Uniquenesses \\
  \hline
    Neat $-$ Scruffy & 0.127 & {\bf 0.491} & 0.145 & 0.502 \\
    Filthy $-$ Clean & $-$0.149 & $-$0.174 & $-$0.194 & 0.666 \\
    Disgusting $-$ Delightful & $-$0.326 & $-$0.145 & $-$0.156 & 0.622 \\
    Beautiful $-$ Ugly & 0.286 & 0.166 & 0.266 & 0.689 \\
    Cool $-$ Youthful & $-$0.383 & 0.374 & $-$0.007 & 0.811 \\
    Sophisticated $-$ Na\"ive & 0.092 & 0.299 & 0.355 & 0.731 \\
    (Long $-$ Short) Hair & {\bf 0.410}  & 0.071 & 0.010  & 0.970 \\
    (White $-$ Brown) Skin & {\bf 0.432} & 0.126 & 0.139 & 0.667 \\
    Short $-$ Tall & 0.089 & $-$0.194 & $-$0.115 & 0.722 \\
    Weak $-$ Strong & 0.018 & $-$0.175 & $-$0.265 & 0.786 \\
    Wan $-$ Robust & 0.311 & $-$0.065 & $-$0.254 & 1.018 \\
    Cautious $-$ Brave & 0.204 & 0.040  & $-$0.291 & 0.599 \\
    Unambitious $-$ Ambitious & 0.150  & $-$0.025 & {\bf $-$0.518} & 0.641 \\
    Masculine $-$ Feminine & {\bf $-$0.422} & 0.277 & {\bf 0.438} & 0.775 \\
    Fulfilling $-$ Empty & 0.119 & 0.263 & 0.473 & 0.492 \\
    Happy $-$ Unhappy & 0.382 & 0.337 & 0.103 & 0.477 \\
    Soft $-$ Firm & 0.158 & $-$0.250 & 0.288 & 0.759 \\
    Elegant $-$ Ungracious & 0.318 & 0.379 & $-$0.041 & 0.544 \\
    Lazy $-$ Hardworking & $-$0.035 & {\bf $-$0.441} & $-$0.194 & 0.758 \\
    Incorrect $-$ Correct & $-$0.010 & $-$0.199 & 0.057 & 0.849 \\
    Sensitive $-$ Insensitive & 0.051 & {\bf 0.480}  & 0.188 & 0.622 \\
    Simple $-$ Complex & 0.292 & $-$0.080 & $-$0.062 & 0.856 \\
    New $-$ Old & 0.024 & $-$0.029 & {\bf 0.566} & 0.606 \\
    Disorganized $-$ Organized & $-$0.067 & {\bf $-$0.553} & 0.178 & 0.529 \\
    Stubborn $-$ Flexible & {\bf $-$0.419} & 0.227 & 0.142 & 0.763 \\
    Closed  $-$ Open$-$Minded & $-$0.23 & 0.209 & {\bf $-$0.488} & 0.751 \\
    Unsocial $-$ Social & $-$0.246 & 0.300   & $-$0.337 & 0.869 \\
    Unfriendly $-$ Friendly & $-$0.073 & 0.233 & 0.216 & 0.808 \\
    Emotional $-$ Intelligent & 0.004 & $-$0.372 & 0.345 & 0.755 \\
    Forgetful $-$ Long-Memoried & 0.171 & $-$0.159 & $-$0.245 & 0.805 \\
    Incompetent $-$ Competent & 0.151 & $-$0.300  & $-$0.328 & 0.609 \\
    Individual $-$ Characterless & 0.017 & 0.100   & {\bf 0.564} & 0.653 \\
    Walking Dictionary $-$ Ignorant & $-$0.127 & {\bf 0.412} & 0.332 & 0.457 \\
    Deep $-$ Shallow & 0.088 & 0.399 & 0.198 & 0.634 \\
    Popular $-$ Unpopular & 0.351 & 0.338 & 0.156 & 0.436 \\
    (Similar to $-$ Different from) Myself & {\bf 0.576} & 0.049 & $-$0.076 & 0.769 \\
    Worthy $-$ Unworthy & 0.226 & 0.382 & 0.207 & 0.400 \\
      \hline
\end{tabular}
\end{center}
\end{table}

}
{
\baselineskip=6mm
\newpage
\bibliographystyle{ECA_jasa} 
\bibliography{paper-ref} 
\end{document}